# Polarization dependent spike formation on black silicon via ultrafast laser structuring

Skoulas E.[a,b], Mimidis, A[a,b], Demeridoy I.[a,c], Tsibidis GD [a,\*], Stratakis E.[a,c,\*]

[a]Institute of Electronic Structure and Laser (IESL), Foundation for Research and Technology (FORTH), N. Plastira 100, Vassilika Vouton, 70013, Heraklion, Crete, Greece
[b]Materials Science and Technology Department, University of Crete, 71003 Heraklion, Greece
[c]Department of Physics, University of Crete, 71003 Heraklion-Crete, Greece

**Abstract**
A comparative experimental and theoretical investigation is presented that centres on the effects of structuring black silicon surfaces with linearly, circularly and azimuthally polarized laser pulses under $SF_6$ ambient atmosphere. It is shown that the asymmetric elliptical micro-cone formations induced by linearly polarized beams result in variable light absorption due to their spatial asymmetry. By contrast, the use of azimuthally polarized beams leads to an omni-directionality of the elliptical cone orientation which is dependent on the local electric field during laser scanning. The locally variant electric field state that is azimuthally polarized leads to a selective conical orientation for the induced structures. The omni-directional conical distribution induced by azimuthal polarization produces similar, angle-independent absorption in the visible spectrum with the symmetrical conical structures that could only be realized with circularly polarized beams.

## 1. Introduction

Silicon is one of the most abundant and significant materials of the 21[th] century. A textured silicon surface is highly absorbing on a broad spectral range extending from the visible to some part of the mid infrared spectra and it is usually termed as black silicon [1]. It is a well-known material that can be produced with various techniques [2]. Almost all daily devices contain parts of processed crystalline silicon as it is a very important material for a wide range of applications ranging from electronics to solar cell technologies and light sensing [3]–[6]. Ultrafast pulsed lasers were first utilized for the black silicon production by Her. et al [7]. Since then, laser structured black silicon surfaces are widely used as optoelectronic response surfaces with impressive light absorbing, water repellent, biological and antibacterial properties [8]–[12]. Structuring silicon surfaces with ultrashort laser pulses in controlled $SF_6$ atmosphere is one of the most popular methods to create black silicon as it is a relatively simple and fast technique compared to the use of chemistry-based approaches.

Interestingly, the addition of reactive $SF_6$ gas during laser processing leads to the formation of structures with sharper features compared to patterning in atmosphere [2], [13], [14].

Among the various types of surface structures that can be fabricated on black silicon, conical shaped structures is one type that has been related with the demonstration of remarkable properties [15], [16]. Analogous properties are also exhibited by various species in nature due to the fact that their surface is covered with similar structures [17], [18]. More specifically, the surface of lotus leaf is patterned with such structures that yield impressive wetting properties [19] and therefore, transferring the natural paradigm in artificial materials is an intriguing task. Extensive studies have also showed that black silicon can be used for the fabrication of neuronal scaffolds, due to the ability to direct the neuron path according to the surface morphological characteristics [20]. In other applications, black silicon is used as a strong absorber for sensitive night vision camera sensors, spectrometers and photovoltaics [21]–[23].

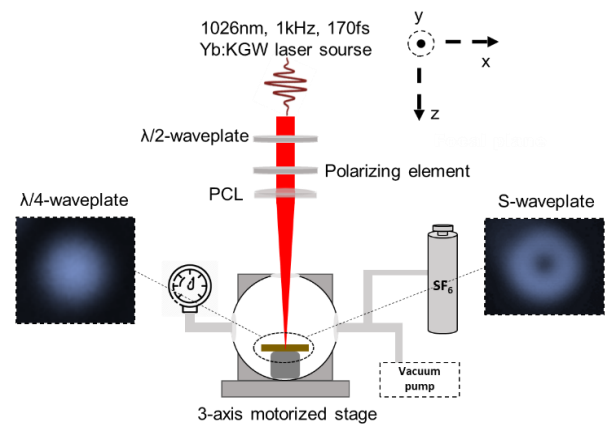

*Fig. 1. Experimental set up for the fabrication of black silicon with linear, radial and circular polarization states in controlled $SF_6$ environment. The polarizing element can be either a λ/4 waveplate or an s-waveplate. PCL stands for a plano convex lens and the images of the beam profile close to the focal plane for Gaussian and donut shaped beams are depicted*

In regard to the laser-based patterning of black silicon with ultrashort linearly polarized pulses, it was shown that conical structures are induced with an elliptically shaped base with a major axis dependent on the polarization of the laser beam [24]. By contrast, texturing with circularly polarized pulses changes the shape of the base of the conical structure to a more symmetrical one [25]. Therefore, it appears that the laser polarization is very crucial for the surface morphological characteristics and the properties of the patterned material. Texturing of metallic, semi-conductive and dielectric surfaces with complex polarization states and production of morphologies of enhanced complexity has been demonstrated in several past reports (for static [26]–[29] or dynamic irradiation [30], [31]). However, to the best of our knowledge, there is no previous study that focuses on the laser induced conical formation with the use of spatially variant polarization states. The aim of the present work is to highlight the important effect of the laser polarization state on the shape of the induced conical structures with the use of azimuthal polarization state. The investigation was carried out for texturing black silicon with three different polarization states, linear, azimuthal and circular. Femtosecond laser pulses were used to acquire black silicon surfaces decorated with conical structures on *p*-doped crystalline silicon surfaces under 0.65 bar of $SF_6$ environmental atmosphere. Differences of the spatial features of the induced patterns are discussed both experimentally and theoretically, while a detailed analysis of the morphological features was followed to compare them with the optical properties of the patterned surfaces. With the utilization of beams with azimuthal polarization, the micro cones' orientation can be controlled while maintaining their elliptical shape leading to a large area omnidirectional conical pattern. Furthermore, the effect of the laser polarization is examined with absorbance measurements highlighting the significance of the conical shape with respect to the angle-dependent absorption efficiency. The omnidirectional elliptical cones mitigate the absorbance anisotropy stemming from the cones' asymmetrical shape.

## 2. Experimental protocol

Laser processing is performed by an Yb:KGW Pharos – SP laser system from Light Conversion that was emitting at 1026 nm central wavelength with 1kHz repetition rate and 170 fs pulse width. The samples were *p*-doped, 1mm thick crystalline silicon at [110] crystal orientation. The samples were placed on a motorized 3-axis stage holding a vacuum chamber connected with a bottle of $SF_6$ as showed at Fig. 1. The vacuum value achieved on the chamber was $10^{-3}$ Pa before the $SF_6$ was released and eventually reach 0.65 bar. The laser beam was guided through the chamber through silver mirrors and focused on the sample surface with an $f = 200$ mm plano convex lens and all irradiations were performed at normal incidence. The spot size radius was characterized from a CCD camera close to the focal plane and was estimated around ~60 μm for Gaussian and ~67 μm for the cylindrical vector beam. Irradiation was performed within the Rayleigh range of the focal position and the number of pulses receptive to the sample for static irradiations were controlled with an external mechanical shutter. The peak fluence ($\Phi$) values were calculated for the Gaussian and the CV beam extracting the zero-intensity null in the center of the annular spot which was ~9 μm in diameter. The peak fluence estimation was performed according to [32] for the Gaussian and [33] for the CV beam. All surfaces were produced with 0.5mm/s velocity and with varying line separation distance in order to preserve the effective number of pulses $N_{eff}$ ~ 200 receptive to the surface the same for all polarization types.

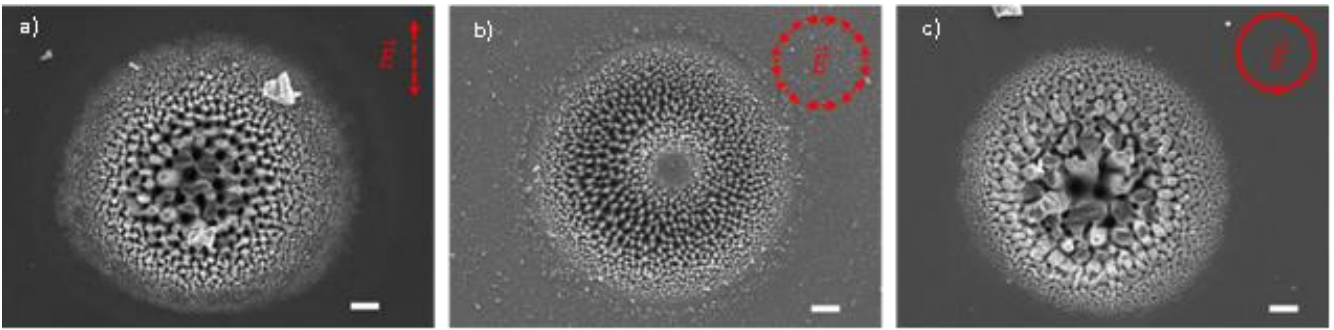

*Fig. 2. Multiple pulse irradiations of 200 pulses at $\Phi = 0.38$ J/cm$^2$ for a) Linear, $\Phi = 0.32$ J/cm$^2$ for b) Azimuthal and $\Phi = 0.38$ J/cm$^2$ for c) Circular polarization states. The white scale-bar is equal to 10 μm. Note the small circular area of non-irradiations of 200 pulses at $\Phi = 0.38$ J/cm$^2$ for a) Linear, $\Phi = 0.32$ J/cm$^2$ for b) Azimuthal and $\Phi = 0.38$ J/cm$^2$ for c) Circular polarization states. The white scale-bar is equal to 10 μm. Note the small circular area of non-irradiated material surface which corresponds to a doughnut shaped beam in b)*

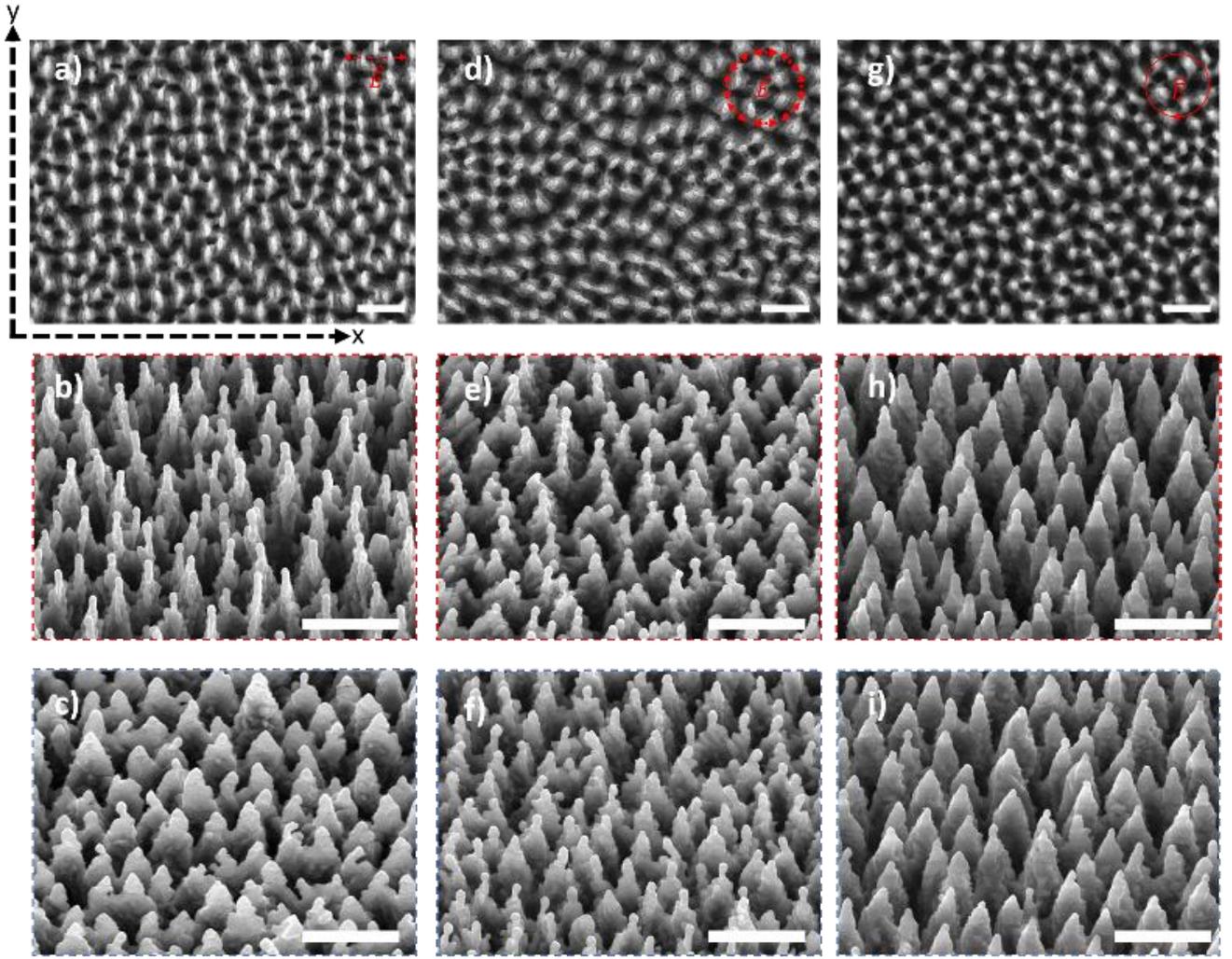

*Fig. 3. SEM images of spike formation with Neff ~200 pulses. Top view SEM images produced with a) Linear, b) Azimuthal, c) Circular polarization, images d) and g) are tilted at $45^0$ perspective of image a) viewing from x axis for d) and y axis for g) respectively. Images e) and h) are tilted at $45^0$ perspective of image b) viewing from x axis for e) and y axis for h) respectively. Images f) and i) are tilted at $45^0$ perspective of image c) viewing from x axis for f) and y axis for i) respectively. The white scale-bar is equal to 10 μm*

The laser power was modulated from the laser amplifier settings and the polarization was controlled with a half wave plate which took place before a polarizing element as depicted at Fig. 1, the polarizing element was for the circular polarization a quarter waveplate and an s-waveplate for achieving cylindrical vector beam with azimuthal polarization. At Fig. 1 the beam profile close to the focal place is depicted for each polarizing element either λ/4 or s-waveplate. In order to acquire the absorption measurements a UV-VIS lambda 950 spectrometer with an integration sphere was used to measure reflectance and transmittance of the black silicon surfaces at two probes with ~$10^0$ angle of incidence from 400 nm up to 2000 nm with 2 nm interval for the reflectance measurement.

The size of the geometrical features of the textured surfaces were calculated from the acquired top view and tilted SEM images for every polarization type. The calculation of the height and width values is based on the average value of at least ten individual measurements to take into account their size variation. The periodicity is computed through a two-dimensional fast Fourier transformation (2D-FFT) on the respective top-view SEM micrographs. Furthermore, the periodicity and the width of the structures was measured in both vertical and horizontal dimensions.

## 3. Results and discussion

The geometric profile of the spike formation on the silicon surfaces was initially investigated through a parametric study, following static irradiation by a variable number of pulses and peak fluences for all three polarization states. Images from scanning electron microscopy (SEM) illustrate results following static irradiation with 200 linearly- (Fig.(2a)), azimuthally- (Fig.(2b)), and circularly- (Fig.(2c)) polarized pulses of $\Phi$ = 0.38 J/cm$^2$, 0.32 J/cm$^2$ and 0.38 J/cm$^2$ respectively. Results show that the base of the induced conical structures has an elliptical shape with a major (long) and minor (short) axis on the horizontal plane. Furthermore, it is revealed that the orientation of the long axis of the ellipsis is polarization-dependent with the long axis oriented always perpendicularly to the electric field of the laser beam. This can be more clearly observed on acquired SEM images of large 4 mm$^2$ surfaces covered with spikes after the laser beam raster-scanning.

*Table 1. Summary of geometrical characteristics of the attained spikes and the respective polarization state of the incident beam*

| Polarization state | Periodicity (μm) | Average width (μm) | Average height (μm) | Density (cm$^{-2}$) |
|---|---|---|---|---|
| Linear Horizontal (x-axis) | 4.1 ± 1.1 | 1.0 ± 0.3 | 2.6 ± 0.3 | 4.0 × 10$^6$ |
| Linear Vertical (y-axis) | 5.5 ± 4.5 | 3.1 ± 0.4 | | |
| Azimuthal Horizontal (x-axis) | 5.6 ± 2.0 | 2.1 ± 0.6 | 2.9 ± 0.5 | 3.6 × 10$^6$ |
| Azimuthal Vertical (y-axis) | 6.4 ± 1.7 | 1.3 ± 0.4 | | |
| Circular Horizontal (x-axis) | 5.1 ± 2.9 | 1.8 ± 0.4 | 3.6 ± 0.4 | 4.6 × 10$^6$ |
| Circular Vertical (y-axis) | 5.0 ± 2.6 | 1.7 ± 0.3 | | |

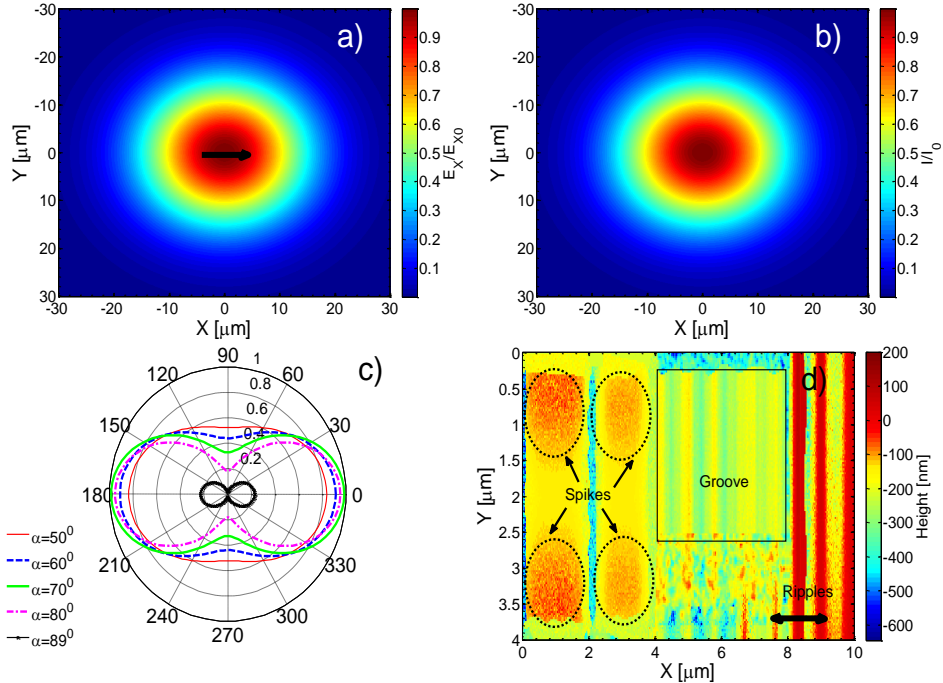

*Fig. 4. (a) x-polarized fundamental Gaussian mode, (b) Intensity profile for linearly polarized beam, (c) Absorptivity as a function of β and the incident angle α. (d) Simulations for surface modification of Si (upper view) [31]*

Fig. 3 illustrates SEM images for each polarization state. The large surfaces were created with the exact fluence values used for the laser spots of Fig. 2 and with the effective number of pulses $N_{eff}$ ~ 200 which was calculated for the Gaussian and the CV beam according to [31], [34].

Fig. 3a shows the top view of the reference surface produced with linearly polarized beams. On the other hand, SEM images in Fig. 3b and Fig. 3c are tilted at 45$^0$ in order to observe the spatial distribution and the possibility of the presence of morphological differences for two-observing axis. Results show that the cones are elongated in a fixed orientation which is perpendicular to the *E*-field (y-axis). Although $N_{eff}$ is the same for both laser spots (Fig. 2) and large area scans, the conical formation of the surface at Fig. 3 were more homogenously distributed. The varying topography in the static irradiations of Fig. 2 can be attributed to the energy distribution of the beam profile, resulting to modulated energy deposition across the laser beam imprint. On the contrary this can compensated via laser scanning as the scanning speed and the line separation can distribute the laser energy on the surface for each pulse.

By contrast, at (Fig. 3d, the conical orientation is not fixed, but instead is radially distributed and always vertically aligned to the existing local electric field which, in this case, corresponds to azimuthal polarization. While, the tilted SEM images Fig. 3e and Fig. 3f indicate that there does not exist any significant change of the spike spatial characteristics that depends on the observation angle due to the fact that there were both high and low aspect ratio spikes which appear equally regardless of the observation angle. Their thickness along the x and y directions are $(2.1 \pm 0.6)$ μm and $(1.3 \pm 0.4)$ μm respectively. On the other hand, Fig. 3g shows that the conical shape of the structures induced by using circular polarization are symmetric, with the long and short axis being almost equal. This indicates that no preferential orientation is exhibited. These results reveal that orientation of the laser induced spikes can be fully controlled through the modulation of the laser polarization state. To further probe the morphological features of the textured surfaces for the three polarization states, optical energy absorption is evaluated on the produced surfaces for two orthogonal angles - probes of incidence. Then, their absorption spectra are comprehended based on their specific geometries which are quantitatively calculated from tilted SEM images in Fig. 3. The sample produced with linear polarization was the reference point for the comparison with the other two polarization states as the absorption measurements were made with the sample tilted at 10 degrees horizontally (probe A) and vertically (probe B) to the preferential elliptical orientation of the cones (i.e. long axis, *y*-axis). Remarkably, in the case of texturing with linearly polarized beams, the obtained conical width distribution is highly asymmetric as overall thickness of the spike varies when observed from the *x* and *y* axis (Fig. 3(a)). This indicates the length difference between the major (direction perpendicular to the electric field of the laser beam) and minor axis of the cross-section ellipse (Fig. 3b,c)). More specifically, in the horizontal direction (along *x*-axis) the thickness (width) of the conical features is $(1.0 \pm 0.3)$ μm while in the vertical direction (along *y*-axis) it is equal to $(3.1 \pm 0.4)$ μm. This approximately threefold increase is attributed to the shape of the microcones which in the vertical direction resembles that of a pyramid. On the contrary, in the vertical direction, their slope becomes significantly steeper and leads eventually to a transition of their shape to that of a cone with much higher aspect ratio. Table 1 summarizes the geometrical characteristics of the micro cones as a function of the respective polarization state which was used to attain them. All the morphological characterization presented in Table 1 was performed on SEM images according to supplemental material. This is confirmed also for azimuthal polarization (Fig. 3(d-f)) for which SEM images show random distribution for each conical structure exhibiting different orientation, although they maintain their elliptical shape. Finally, in the case of circular polarization (Fig. 3(g - i)), results show that the average width remains unchanged between the two perpendicular directions, $(1.8 \pm 0.4)$ μm and $(1.7 \pm 0.3)$ μm, which manifests that the shape of the produced conical base is almost circular.

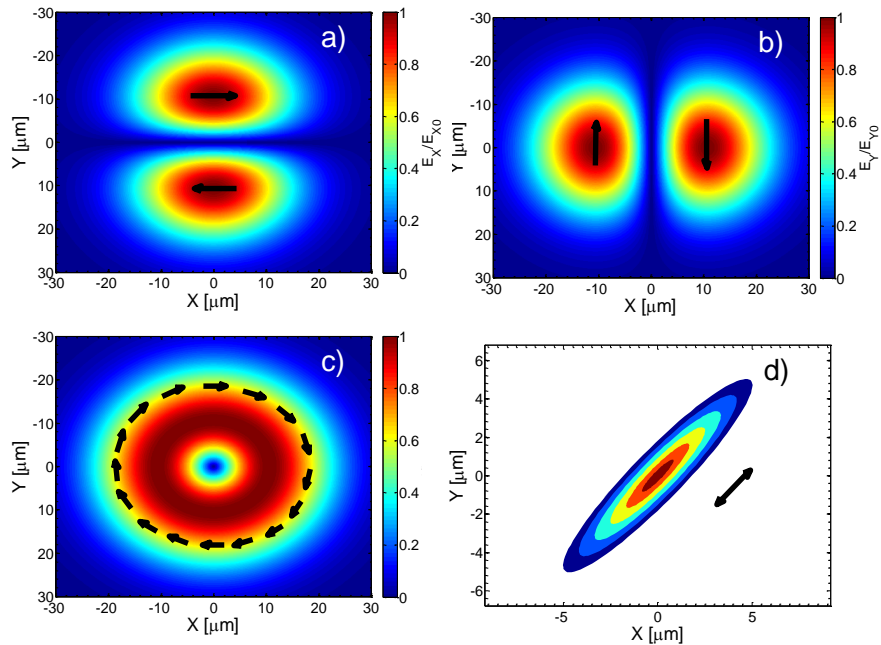

*Fig. 5. x-polarized (a) and y-polarized (b) Hermite-Gauss HG01 modes, (c) Intensity profile for azimuthally polarized beam (black vectors indicate polarization direction), (d) Profile of spike (upper view) (color indicates normalized height)*

Considering the periodicity of the structures, it is evident that the produced structures induced by circularly or azimuthally polarized beams are not characterized by different periodicity in either direction, as evident in Table 1. On the other hand, for linearly polarized beams the periodicity appears to increase along the vertical direction due to the pyramidal shape of the cones which causes a gradual evolution of the profile resembling a ridge. This in the 2D-FFT is perceived as a greater distribution of periodicities affecting the average periodicity and its error value. The density of induced structures remains almost unaffected for every polarization state that was calculated $(4.0 \times 10^6)\ cm^{-2}$ for linear $(3.6 \times 10^6)\ cm^{-2}$ for azimuthal and $(4.6 \times 10^6)\ cm^{-2}$ for circular polarization. Similarly, the average height of the micro cones on the textured surfaces seems to be comparable for all polarization cases. This is mostly attributed to the almost identical peak fluence values used which represents parameter to influence significantly the height of spike [23], [24], [35]. Simulation results in previous reports have shown that the polarization and the produced inhomogeneous energy deposition affects greatly the nanogratings that are induced in materials [36], [37]. More specifically, finite difference time domain simulations show that the electronic modification due to energy absorption and the induced nanogratings are asymmetrical and elongated for linear polarization; by contrast, circularly polarized beams lead to symmetrical electron distribution and therefore the induced structures resemble the ones observed in this work [37]. Similar conclusions can be deduced for azimuthal polarization. While simulations in those reports were limited to the calculation of the electronic distributions, our calculations indicate an elongation perpendicular to the polarization direction for linearly polarized beams.

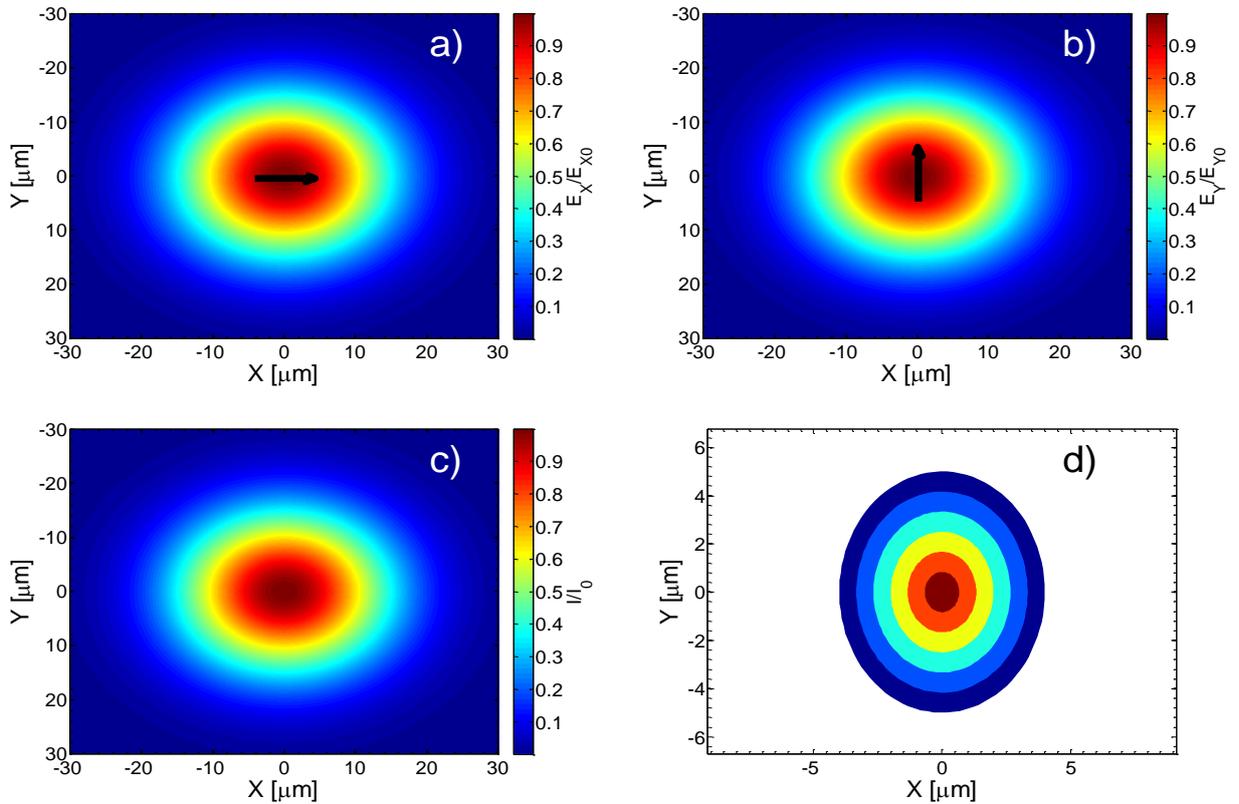

*Fig. 6. x-polarized (a) and y-polarized (b) fundamental Gaussian modes, (c) Intensity profile for circularly polarized beam, (d) Profile of spike (upper view) (color indicates normalized height)*

The simulation details for the formation of spikes/cones have been described in a previous study that showed that grooves (supra-wavelength structures parallel to the polarization direction) constitutes an intermediate state before the production of cone-like structures [38]. This is the first attempt to explain the characteristics (i.e. size along *x*-axis and y-axis) of the protruded structures in a multiscale modelling approach that involves phase transition to describe the fluid dynamics and resolidification that leads to final surface profile. Certainly, a similar approach towards computing the morphological details of spikes for azimuthal and circular polarization should also be performed to deduce a more accurate polarization-dependent shape. Nevertheless, the focus of simulations in this work is centered on the correlation of elongation of spikes with energy absorption; more specifically, the difference in the energy absorption which is dictated for the three the polarization states is expected to enhance material removal along particular directions. To quantify the response of the material to specific laser conditions, the absorbed energy is correlated with the resultant intensities for the component electric fields along the *x*- and *y*-axis. The electric field components along *x*- and *y*-directions as well as the intensity profiles (normalized values) are illustrated for linearly (Fig. 4), azimuthally (Fig. 5) and circularly polarized (Fig. 6) beams. For linearly as well as circularly polarized beams that can be regarded as superposition of linearly polarized beams, linearly polarized Gaussian modes are used. By contrast, for azimuthal polarization, *x*-polarized and *y*-polarized $HG_{01}$ Hermite-Gauss modes are used [39]. The objective of the simulations is to deduce a simple derivation of the features of the induced spikes based on the energy absorption. More specifically, to compute the effect of the incident (plane polarized) beam that irradiates a surface covered with spikes at an angle $\theta$, [40], the energy absorptivity is computed through the expression $A(\beta) = A_{(s)}(\beta) \sin^2(\alpha) + A_{(p)}(\beta) \cos^2(\alpha)$ where $\alpha$ corresponds to the angle between the polarization direction and the plane of incidence; on the hand $\beta$ stands for the angle incidence of the laser beam on the spike surface. By contrast, for circularly polarized beams, the above expression takes the form $A(\beta) = (A_{(s)}(\beta) + A_{(p)}(\beta))/2$ where $A_{(s)}(\beta)$ and $A_{(p)}(\beta)$ are Fresnel formula for $s$- and $p$-absorption [25], [28], [41]

$$A_s(\beta) = \left| \frac{\cos(\beta) - \sqrt{n^2 - \sin^2(\beta)}}{\cos(\beta) + \sqrt{n^2 - \sin^2(\beta)}} \right|^2 \quad (1)$$

$$A_p(\beta) = \left| \frac{-n^2(\beta) + \sqrt{n^2 - \sin^2(\beta)}}{n^2 \cos(\beta) + \sqrt{n^2 - \sin^2(\beta)}} \right|^2$$

The calculation of the absorbed energy requires also the knowledge of the refractive index $n$ and extinction coefficient $k$. ($n = 3.5662$ and $k = 0.00027004$ for laser wavelength $\lambda_L = 1.026$ nm in the unexcited state) [42]. Although, excited material exhibits variable optical properties [43] and therefore energy deposition changes, the spatial distribution of the laser energy absorption always exhibits the same monotonicity and therefore, for the sake of simplicity and to estimate the absorptivity only the optical properties in the unexcited state are considered in the calculations.

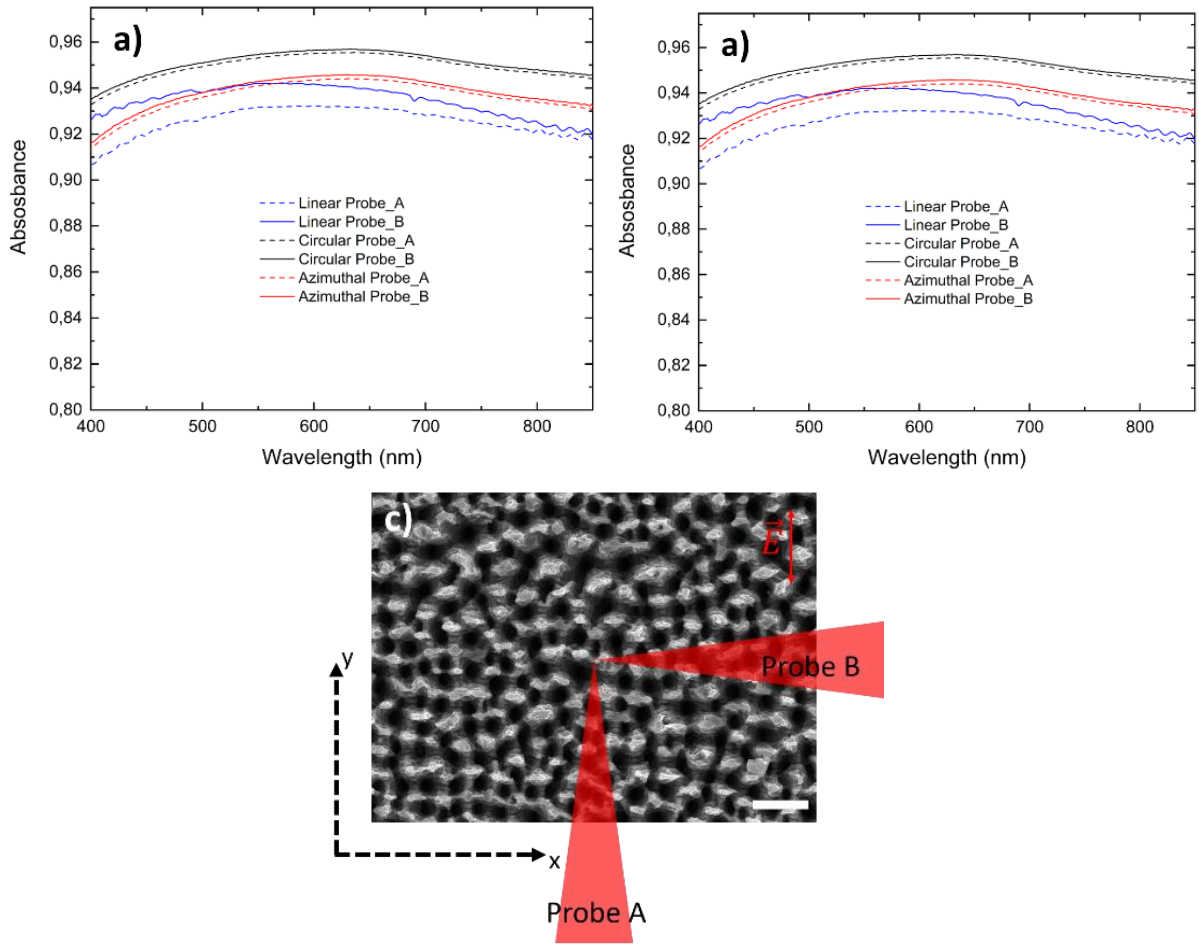

Fig. 7. Absorption spectra for (a) VIS and (b) IR region for surface produced with linear (blue), radial (red) and circular (black) polarization. The dashed lines represent the configuration where the sample is placed according to probe while the continuous line on probe B. SEM image (c) on the right indicates UV-VIS light incidence of ~ 10 degree angle with respect of the spike elipticity produced with linear polarization

Theoretical calculations have been conducted for five representative values of the angle of incidence $\alpha$ in the range $[50^0, 89^0]$ for $\beta \in [0, 2\pi]$ (Fig. 4c). It is evident that enhanced energy absorption is predicted for $\beta$ close to 0 or $2\pi$; this angle for laser conditions that lead to ablation, induces structures which appear to be elliptical with the longer axis being perpendicular to the polarization vector of the beam. This increased energy absorption is more pronounced at increasing angle of incidence (Fig. 4c) that leads to a significant decrease of the size of the minor axis of the ellipse. Multiscale simulations performed in previous reports confirm conclusively the formation of such structures (Fig. 4d). Similar theoretical predictions can be derived for azimuthally polarized beams (Fig. 5) while the use of the expression $A(\beta) = A(\beta) + A(\beta))/2$ which is independent of $\alpha$ shows a symmetric distribution that leads to spikes with an almost circular base (Fig. 6).

The impact of the aforementioned surface patterns on the optical properties has also been analyzed by evaluating the corresponding absorption spectra for every polarization state. The samples were placed with two different configurations (probe A and B) with respect to the plane of incidence of the UV-VIS measuring rays. More specifically, on probe A, the plane of incidence is parallel to the semimajor axis of the elliptic cross-section of the cones while on probe B the plane of incidence is parallel to the semi-minor axis. In this way, two extreme cases (i.e. parallel and perpendicular to the long axis of the cones) are investigated. Results show (Fig. 8) that light absorption is increased on the surface fabricated with circular polarization.

This effect appears to remain constant for wavelengths spanning the near UV to deep IR region and it is predominantly attributed to the higher aspect ratio of the cones which facilitates a more pronounced light trapping effect [44].

By contrast, the absorption spectra of cones formed after irradiation with linearly and azimuthally polarized pulses are similar with small deviations also arising from the different fluences used during fabrication; the latter is also projected on the different height of the cones (2.6 $\pm$ 0.3) μm for linear and (2.9 $\pm$ 0.5) μm for azimuthal polarization.

As the measurement is changed from probe A to probe B the absorption spectra corresponding to linear and azimuthal polarization diverge more intensively as the reflectivity of the cones increases for probe A for the linear polarization. In the case of azimuthal polarization, a small difference in the absorption values is observed, with the respective absorption for probe A being ~1% higher. This effect can be attributed to fluctuations of the random alignment of spikes. On the contrary, probe A and probe B measurements for the surface produced after irradiation with linear polarization reveals a greater discrepancy between respective spectra, with probe B absorbing > 1.5% over the whole region of the electromagnetic spectrum investigated and > 3% at longer wavelengths. This effect is attributed to the varying slope of the cones depending on axis of the measurement (Fig. 3). The difference in width between the two configurations eventually leads to a significant variation to the respective aspect ratio given the produced structures have the same height. This affects the light trapping effect which is more pronounced for configurations of higher aspect ratio. Finally, the spectra corresponding to circular polarization show that the aspect ratio of the microcones produced with this polarization state remains unaffected between different configurations. For intermediate angles between the two extreme configurations of probe A and B, we expect the absorption spectra to be contained between the respective probe A and B spectra.

## 4. Conclusions

In conclusion, a comparative study was presented in which the influence of the laser beam polarization on the shape of conical structures induced on black silicon induced with ultrashort laser pulses was highlighted. Special emphasis was placed on the omnidirectional spike formation with azimuthally polarized laser pulses. Results indicate that the shape of the conical structures is directly related to distinct optical properties. The distinct optical properties due to the different conical cross section and the capability to control the conical shape via modulating the beam polarization could potentially constitute an important tool for selective light absorbers or sensors depending on the angle of light incidence.

**Acknowledgements**

The authors acknowledge financial support from Greece and EU-ESF Fund through HRDELL 2014-2020 program (project MIS 5004385). The research was co-financed by Greece and European Union (European Social Fund-ESF) through the Operational Program "Human Resources Development, Education and Lifelong Learning 2014-2020" in the context of the project 'Formation of biomimetic micro/nano structures via polarization modulation of ultrashort-pulsed lasers' (MIS 5004385).

Corresponding authors: tsibidis@iesl.forth.gr; stratak@iesl.forth.gr